\documentclass[%
 reprint,
 amsmath,amssymb,
 aps,
]{revtex4-2}

\usepackage{graphicx}
\usepackage{dcolumn}
\usepackage{bm}
\usepackage{nomencl}
\makenomenclature
\graphicspath{ {./figure/} }

\begin{document}

\title{Viscoelastic material properties determine contact mechanics of hydrogel spheres}

\author{Chandan Shakya}
 \affiliation{Van der Waals-Zeeman Institute, Institute of Physics, University of Amsterdam, Amsterdam, The Netherlands \\$^{2}$Physical Chemistry and Soft Matter, Wageningen University, Wageningen, Netherlands}
 \email{c.shakya@uva.nl}

\author{Jasper van der Gucht}
 \affiliation{Physical Chemistry and Soft Matter, Wageningen University, Wageningen, Netherlands.}
 \email{jasper.vandergucht@wur.nl}
 
 \author{Joshua A. Dijksman}
 \affiliation{Van der Waals-Zeeman Institute, Institute of Physics, University of Amsterdam, Amsterdam, The Netherlands \\$^{2}$Physical Chemistry and Soft Matter, Wageningen University, Wageningen, Netherlands}
 \email{j.a.dijksman@uva.nl}

\date{\today}

\begin{abstract}
Granular materials are ubiquitous in nature and industry; their mechanical behavior has been of academic and engineering interest for centuries. One of the reasons for their rather complex mechanical behavior is that stresses exerted on a granular material propagate only through contacts between the grains. These contacts can change as the packing evolves. This makes any deformation and mechanical response from a granular packing a function of the nature of contacts between the grains and the material response of the material the grains are made of. We present a study in which we isolate the role of the grain material in the contact forces acting between two particles sliding past each other. We use hydrogel particles and find that a viscoelastic material model, in which the shear modulus decays with time, coupled with a simple Coulomb friction model captures the experimental results. The results suggest that the particle material evolution itself may play a role in the collective behavior of granular materials.

\end{abstract}

\maketitle


\section{\label{introduction}Introduction}

Particle packings composed of soft elements are relevant for many systems, from battery electrolytes to emulsions and biological cells. These particle packings are found to behave very differently from packings primarily made up of rigid particles. For instance, deformable particle packings will pack to a higher packing fraction than equivalent packings made up of rigid particles as found already in the classic work on pea packings by Weaire et. al.\cite{weaire2008pursuit} as well as in recent 2D simulations by Cardenas et. al.\cite{cardenas2020compaction}. The flow properties of soft particle packings are also different from their rigid particle counterparts as shown in studies by van der Vaart et. al. and Campbell et. al. \cite{van2013rheology,campbell2006granular}. To better understand soft particle packings, it is instructive to understand how one particle interacts with another via a single contact. After all, for a packing to undergo any bulk deformation either the constitutive grains must squeeze or extend, or the constitutive particles need to rearrange. For any sort of rearrangement to occur, at least one of the particles must slide past at least one of the others, as the particles deform. Additionally, in packings made of grains, under any kind of slow, quasi-static load application, force balance on particles is maintained at all times, via an appropriate redistribution of contact forces during the structural deformation in the packing. Any time dependence in the contact mechanics of the material or the contacts disrupts such force balance. The relaxation of bulk material properties that is so typical for soft materials can so give rise to spontaneous force imbalance situations in static packings made of soft grains. Soft granular packings can so experience particle motion, resulting in macroscopically observable (slow) change in strain, such as creep. In this mechanism, the observed creep would occur without any change in grain's arrangement or fracture of the grains constituting the packing, which is not a mechanism that have been often associated with granular creep before in studies such as \cite{ando2019peek,vandamme2009nanogranular,lade1998experimental,leung1997role}, that mostly focus on harder grains. Particle interactions studies have been conducted before such as the ones by Nardelli. et. al. \cite{nardelli2019experimental} in a study of the evolution of normal and tangential forces as particles slid past each other. However, this study was limited to relatively hard grains and the effect of deformability could not be studied in this experiment. Tsai et. al.\cite{tsai2020force} studied the evolution of tangential and normal loading for deformable polydimethylsiloxane (PDMS) spheres immersed in a mixture of glycerol and water for sphere passing and fixed depth experiments. Giustiniani et. al. \cite{giustiniani2018skinny} studied the deformation evolution of two droplets enclosed in an elastic adhesive film in shear and found a time-dependent exponential decay relationship for the angle between the surface of the drop and the needle holding the drop. This study also provides insight into why such a relationship exists for such a system and compares the deformation in the experiment with a surface tension-driven simulation. Louf et. al.\cite{louf2021poroelastic} also tracked the relaxation of the shape of a hydrogel bead  in a bed of harder spheres and was able to find a timescale associated with it. The present study describes the experimental force response during a sliding event for a pair of initially spherical particles capable of significant deformation. Specifically, the study examines the evolution of horizontal and vertical forces of low friction, highly deformable particles, immersed in water with undetectably small interparticle adhesion.

\section{\label{setup_description}Description of the experiment}
The evolution of forces during contact formation and contact loss between a pair of soft particle spheres was investigated using two custom setups. A plate-plate compression setup is used to look at the force response of a particle as a function of time at a fixed overlap. In this case, the particles were loaded normally (i.e. center-to-center), and the evolution of the normal force was measured as a function of time. The custom inter-particle shear setup is used to investigate the change in contact forces as particles slide past each other, as in the case of particle rearrangement in a packing. In this setting, the particles are loaded obliquely (i.e. there is shear and compression) and both components of the response forces are expected to change in magnitude. The model material used in this experiment were polyacrylamide hydrogel spheres produced by Educational Innovations Inc (GB-710). The setups used are custom-made and described in the following sections.


\subsection{\label{shell}Sample preparation}
The hydrogels are grown in milli-Q water. They typically have a diameter of 18.5 mm after swelling. The hydrogel spheres to be tested are half-dyed with Nile blue perchlorate and their dimensions perpendicular to the dye front are measured. Measurements of the hydrogel sphere diameter are taken 3 times. The gels are then placed with the dye front visible on the shell to be described below and fastened in place with the collar. The dimensions from the bottom of the shell to the tip of the gel sphere are measured with Vernier calipers.

The sample (in this case the hydrogel sphere) is placed in a custom 3D-printed geometry. The geometry was inspired by the container devised by Tsai et. al.\cite{tsai2020force}. The geometry consists of a cylindrical block with a hemispherical cutout (shell) that fits the hydrogel sphere to be tested.  The spheres are held in place with a collar that has a hemispherical cutout matching that of the shell. The hydrogel sphere is held in position by screwing the collar to the block. Measurements for 3D printed geometry are provided via a 2D drawing in a supplementary autocad file.

\subsection{\label{Inter_particle_shear} Inter particle shear}

\begin{figure*}
\includegraphics[scale=0.35]{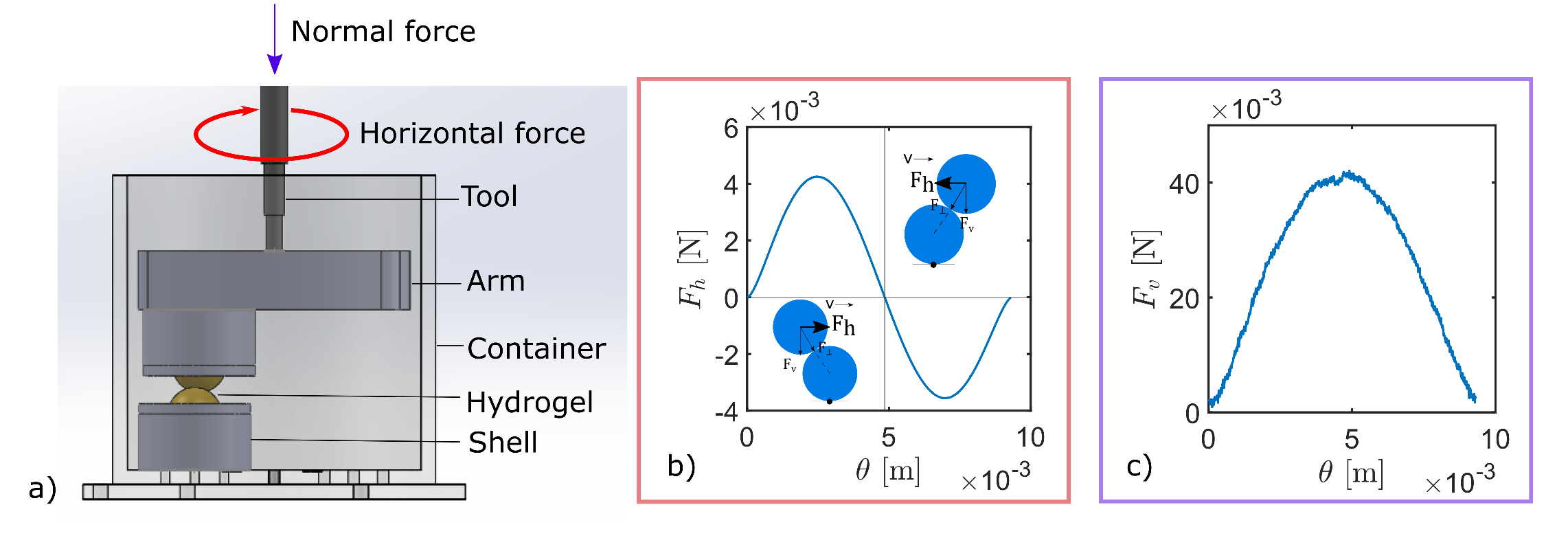}
\caption{\label{fig:setup_shear} a) Experimental Setup: Inter-particle shear. b)Forces in the horizontal direction $(F_h)$ vs. linear angular displacement $(\theta)$  in the rotational direction. As the moving particle approaches the stationary particle as shown by the diagram on the bottom left, the $F_h$ acts towards the right, which is defined as the positive force. As the moving particles moves away from the stationary particle the $F_h$ acts towards the left, which consequently gives a negative force. c)Forces in the vertical direction $(F_v)$ vs. linear angular displacement $(\theta)$ in the rotational direction}
\end{figure*}

The inter-particle shear setup was built based on previous work by Workamp et. al and Rudge et. al. \cite{workamp2019contact,rudge2020natural}. A schematic of the setup used in the set of experiments described in this experiment is shown in figure \ref{fig:setup_shear}a). A custom 3D cylindrical container is placed on the rheometer (Anton Paar MCR 501). A shell described in section \ref{shell} is screwed in place to the cylindrical container. Since the container is 3D printed and the spots to screw in the shells are pre-determined, this fixes the position of the shell relative to the center of the cylinder and the rheometer tool. Another shell is screwed to a rotating arm, which can be connected to the rheometer. The distance between the spots to screw the fixed shell and the axis of rotation was set to 27 mm in the 3D model of the rotating arm to ensure that the shell in the container and the shell in the rotating arm slide over each other. The dimensions of the rotating arm and the cylinder, along with the position of the screws, are also provided in the 2D autocad file attached with this paper.

\paragraph*{Protocol}

The cylindrical container is placed on the rheometer and a zero gap is performed with the custom axis fixed to the rotating arm. The tool and rotating arm are then moved up and removed. The half-dyed hydrogel spheres are then placed into their respective shells and the collar is screwed into the shell. To test for a snug fit, we manually try and rotate the hydrogel in the shell and collar assembly. If the hydrogel sphere rotates in the shell, a larger half dyed sphere is used. We attempt to rotate the new hydrogel in the shell as well. The process is repeated until we find a hydrogel sphere large enough such that it does not rotate in the shell-collar while manually tested. The manual rotation tests involve much larger forces than the rheometric test and so establish that no rotation of the fixed spheres occurs. Then the shells with the spheres can be fixed onto the container and the rotating arm and placed back into the measuring system. For the measurements shown in this study, the diameter of the upper hydrogel sphere is $18.4 \pm 0.195$ mm and the diameter of the lower hydrogel sphere is $19.1 \pm 0.265$ mm, as established by caliper measurements on different sides of the used spheres. The height of the the shell and the upper hydrogel sphere together is 29.53 mm and the height of the shell with the lower hydrogel sphere is 31.25 mm. Thus, the tool is brought down to a gap of 65mm to ensure the particles are not in contact. Milli-Q water is poured into the container until both the shells and the rotating arm are completely immersed. 

The tool is then lowered 1 mm at a time and manually rotated to check for contact. Once contact is detected by observing a finite torque needed to rotate the tool, the probe is moved 1 mm up and locked at the instrument's deflection angle reference position.

The tool is first rotated at 0.5 rpm. During this rotation, when a torque of 0.2 mN.m is detected, the rotating arm is forced to stop. This torque is small enough to detect contact but larger than the initial torque that the rheometer has to apply in order to start the experiment from rest. If the specified torque level is not detected, the tool will move down 0.1 mm. This procedure was carried out to obtain an estimate of where the contact would occur in terms of displacement. Once the displacement required to obtain contact is known, further testing could be done using this displacement as a reference point. At low rotational velocities this would mean we would only need to conduct the test between known displacements, avoiding extraneous movement and time. The process is repeated until contact is detected using this procedure. Once contact is detected the moment arm will rotate back to a fixed angle. Then the tool will rotate at a speed of $2.38\times10^{-5}$ m/s ($8\times10^{-4} [rpm]$) for 3000 seconds in the clockwise direction, stay fixed in the final position for 1 minute, rotate back at the same speed for the same duration, and wait another minute. This process is repeated at speeds of $4.75\times10^{-5}$, $1.19\times10^{-4}$ and $2.37\times10{^-4}$ m/s. Once a repetition is completed, the tool is moved down 0.1mm (i.e., the overlap is increased by 0.1 mm). The whole process is repeated at the same set of speeds. A total of five overlap levels, each 0.1 mm apart, were explored. A sample dataset obtained from the measurement is shown in figure \ref{fig:setup_shear}b), which shows the horizontal force response($F_h$) to sliding as a function of the linear angular displacement ($\theta$). Here $\theta$ is the product of angular displacement (in radians) and the length of the moment arm (which is 0.027 m). Figure \ref{fig:setup_shear} c) shows the vertical force response ($F_v$) as a function of $\theta$ during the same experiment.

\subsection{\label{Plate_plate_compression} Plate plate compression}

\begin{figure*}
\includegraphics[scale=0.27]{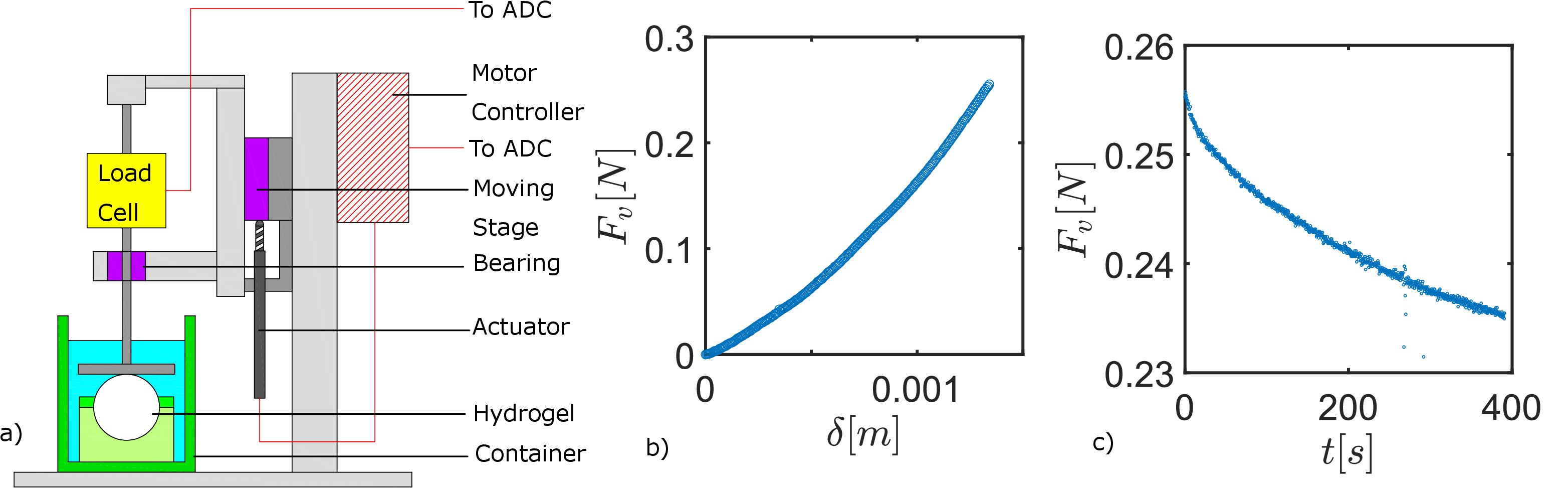}
\caption{\label{fig:setup_compression} a) Experimental Setup: Particle compression. b) Typical Force $(F_v)$ vs. displacement curves $(\delta)$ during indentation c) Typical Force $(F_v)$ vs. time $(t)$ curves when compressive strain is held static}
\end{figure*}

The setup for the plate-plate compression was a modified version of a previously existing custom-built setup used to conduct indentation tests on polymer samples. A schematic of the setup is shown in figure \ref{fig:setup_compression}a). The setup consists of an actuator (Thorlabs Z825BV) connected to a moving stage (Thorlabs MT1) controlled by a motor controller (Kinesis KDC 101) that can interact with a Matlab platform. The moving stage is connected to a metallic rod with a Wheatstone bridge based S-beam load sensor (Futek LSB200 FSH03871). The bridge is supplied with a constant voltage, and the output bridge signal is amplified with an amplifier, both of which were built into a strain gauge input signal conditioner (ICP DAS SG-3016). The amplified signal is then filtered with a low-pass RC filter (10$\Omega$ resistor 10nF capacitor). The filtered signal is then converted to a digital signal with a 14-bit analog-to-digital converter (ADC) present in a National Instrument data acquisition instrument (NI DAQ 6001). The sampling frequency for the load cell was set to 100 Hz. The digital signal can then be read into a computer using a Matlab interface. The sampling frequency for the load cell was set to 1kHz. The load cell is attached to a flat plate that is used to press the sample.  The working of the custom-built setup is also described by Boots et. al.\cite{boots2019development}.

\paragraph*{Protocol}
The shell with the sample is held in place in a larger container with double-sided tape. The container is then filled with water. The compression plate with the load cell is lowered into the container until the entire compression plate is underwater but does not touch the sample to be tested.
An indentation test is performed with the compression plate moving down 0.5 mm at a speed of 0.3 mm/s to find the location of contact. The test results in a displacement ($\delta$) vs vertical force ($F_v$) plot as shown in figure \ref{fig:setup_compression}b). Once contact is found, the compression plate can be retracted and the test is repeated to confirm the location of the contact point.

Once the point of contact is ascertained the compression plate is moved to the location of contact and a strain of 6 percent is applied at a strain rate of 0.002/s. For this experiment the diameter of the hydrogel sphere was $18.63 \pm 0.17$ mm, as established by caliper measurements on different sides of the used sphere. It was compressed by 1.12 mm. The exact velocities and indentation distance are calculated based on the diameter of the particle being tested. Then the compression plate is held static for 3000 seconds. Forces are recorded from the load cell during this entire process at a rate of 100 Hz. After separating the forces during indentation, a typical force relaxation curve is observed as shown in figure \ref{fig:setup_compression}c).

\section{Theoretical Model
\label{theoretical_model}}
Here we explain the models used to analytically describe our experimental results. The following assumptions are applied for the model (i) the particles in contact are perfectly spherical. (ii) the material in the experiment is perfectly homogeneous. (iii) The arc of rotation in the inter-particle shear experiment is sufficiently large as to consider the difference between a linear particle pass and a circular path to be negligible. With these assumptions, the response of the bulk volumetric strain response of the particle is calculated as a sum of two factors; the elastic response (modeled by a Hertzian model) and a dissipative response modeled as a viscoelastic model. There is a long history of considering viscoelastic contact mechanics for spheres, in particular for the use in Discrete Element Methods~\cite{jian2019normal}. For a brief discussion on the history, we refer to Brilliantov et. al. \cite{brilliantov2015dissipative} The combined model used here is based on among others that work and the subsequent work done by Jian et. al. \cite{jian2019normal} which takes a Kelvin-Voigt approach while also taking into account the relaxation effect of viscoelastic materials. Here, for simplicity it is typically assumed that there is one dominant timescale. Briefly, it assumes that the effective shear modulus of a particle is:
\begin{eqnarray}
 G(t) = G_e + \hat{G}(t)
 \\
F_{\perp}(t)=F_e+\hat{F}(t)
 \label{eq:modulus_division}
\end{eqnarray}
\nomenclature{\(t\)}{Time since the start of contact}
\nomenclature{\(G(t)\)}{Shear modulus as a function of time}
\nomenclature{\(G_e\)}{Constant elastic shear modulus}
\nomenclature{\(\hat{G}(t)\)}{Relaxing part of shear modulus which is a function of time}
\nomenclature{\(F_{\perp}(t)\)}{Forces perpendicular to the plane of contact}
\nomenclature{\(F_e\)}{Elastic Forces}
\nomenclature{\(\hat{F}(t)\)}{Dissipative Forces}
Where, $G(t)$ is the total effective shear modulus, $G_e$ and ${\hat{G}}$ are the elastic and relaxable part of the shear modulus and ${\hat{G}}$ is modeled as:
\begin{eqnarray}
\hat{G}(t)=G_l\cdot{e^{-\frac{t}{\tau}}}
\label{eq:modulus_decay_dissipative}
\\
G(t) = G_e + G_l\cdot{e^{-\frac{t}{\tau}}}
\label{eq:modulus_division_expanded}
\end{eqnarray}

$F_{\perp}(t)$ is the net center-to-center force of repulsion between the centers of the particles. $F_e$ and $\hat{F}(t)$ are forces due to the constant elastic part of the shear modulus $G_e$ and the time-dependent dissipative part of the shear modulus, $\hat{G}(t)$. $G_l$ is a prefactor to the exponential function corresponding to the maximum value $\hat{G}(t)$ can take. $\tau$ refers to the relaxation time, characteristic of the material of the grain, and $t$ is simply the time passed since the start of the experiment.

\nomenclature{\(G_l\)}{Maximum relaxable shear modulus}
\nomenclature{\(\tau\)}{Relaxation time}

\subsection{\label{hertz}Hertz's contact model}
 The elastic response resulting from the bulk compression of the particle volume is modeled with a Hertz model:
\begin{eqnarray}
F_e(t) = \frac{{4G_e}\sqrt{R}}{3(1-\nu)} \delta^{3/2}(t),
\label{eq:elastic_equation_force}
\\
G_e=\frac{E^*}{2(1+\nu)};
\\
\frac{1}{E^*} =\frac{1-\nu_1^2}{E_1}+\frac{1-\nu_2^2}{E_2},
\\
\frac{1}{R}=\frac{1}{R_1}+\frac{1}{R_2}
\label{eq:elastic_equation}
\end{eqnarray}
Where $G_e$ is the elastic part of the shear modulus of the system, $R$ is the effective radius, $\delta$ is the overlap between the two interacting spheres in the center-to-center direction, $\nu$ is the Poisson's ratio of the sphere, $E^*$ is the effective elastic modulus of the system, $E_1$ is the Young's modulus of the upper sphere, $E_2$ is the Young's modulus of the lower sphere. In the case of the inter-particle shear experiment, we assume $E_1=E_2=E$ (since the interacting spheres are made of the same material). In the case of the plate-plate compression experiment we assume $E_1=E$ and $E_2=\infty$ assuming the plate stiffness is infinitely larger than the hydrogel sphere stiffness. In reality, the Young's modulus of the hydrogel particle is about 8 kPa and the Young's modulus of the compressing plates and the shells holding the hydrogel is about 2.8 GPa as specified by the data sheet from the 3D printing supplies manufacturer. For both experiments, it is assumed that $\nu_1=\nu_2=\nu=0.5$: hydrogels do not deswell in the context of the modest pressures exerted; the osmotic pressure of the gels is much higher than the local stress Schulze at. al. \cite{schulze2017polymer}.$R_1$ is the radius of the upper sphere and $R_2$ is the radius of the lower sphere. For the inter-particle shear test $R_1=9.2 mm$ and $R_2=9.96 mm$. It is also to be noted that in the case of plate-plate compression, $R_1=\infty$ as the sphere is being compressed with a flat plate and $R_2=9.3 mm$.
\nomenclature{\(E^*\)}{Effective Young's elastic modulus}
\nomenclature{\(E\)}{Young's modulus of sphere}
\nomenclature{\(G_{eff}\)}{Effective shear modulus of system}
\nomenclature{\(\nu\)}{Poisson's ratio of sphere}
\nomenclature{\(R\)}{Effective radius}
\nomenclature{\(R_1\)}{Radius of upper sphere}
\nomenclature{\(R_2\)}{Radius of lower sphere}
\nomenclature{\(\delta\)}{Particle overlap between two spheres}

\subsection{\label{viscous}Viscous Dissipative model}
The time-dependent response seen with the experimental data is modeled using the viscous dissipative model from Jian et. al.\cite{jian2019normal}. Their model calculates the dissipative part of the force response as:
\begin{eqnarray}
\hat{F}(t) = \frac{4\sqrt{R}}{3(1-\nu)} \int_0^t \hat{G}(t-s) \ d[\delta^{3/2}(s)] \,
\label{eq:viscous_force_calculation}
\end{eqnarray}
Where $s$ is the integration factor integrated from the moment of contact $t=0$.

In addition to this model, a cutoff was implemented in the model such that the resultant forces from center to center $F_e$ and $\hat{F}(t)$ were always repulsive or zero to prevent attractive forces, as was observed in all of the experiments performed. The need for this addition is explained in more detail together with the experimental results described in Section \ref{experimental_hz_forces}. Finally, it is assumed that the contact resistance is purely frictional, and thus the classical Coulomb friction model is used to interpret the tangential interactions. It states:
\begin{eqnarray}
F_f(t)=\mu F_{\perp}(t)
\label{eqn:friction}
\end{eqnarray}
Where $F_f(t)$ is the total frictional force tangential to the surfaces in contact and $\mu$ is the coefficient of friction.
\nomenclature{\(F_f(t)\)}{Frictional forces}
\nomenclature{\(\mu\)}{Coefficient of friction}

\section{
\label{experimental_results}
Results and Discussion}

\subsection{\label{interparticle_results}Inter-particle shear experiment results}

In this section, we examine how the time scales affect force responses in the direction of motion and perpendicular to the direction of motion during a particle sliding event. Typical force response in the direction of motion (hereafter, referred to as horizontal direction) and perpendicular to the direction of motion can be seen in figures \ref{fig:setup_shear} b) and c) respectively. The following sections examine the various characteristics to be observed in such data and how the models described in Section \ref{theoretical_model} can help us interpret the experimental results.  

\subsubsection{\label{experimental_hz_forces}Horizontal component of forces}

\begin{figure*}
    \centering
    \includegraphics[scale=0.4]{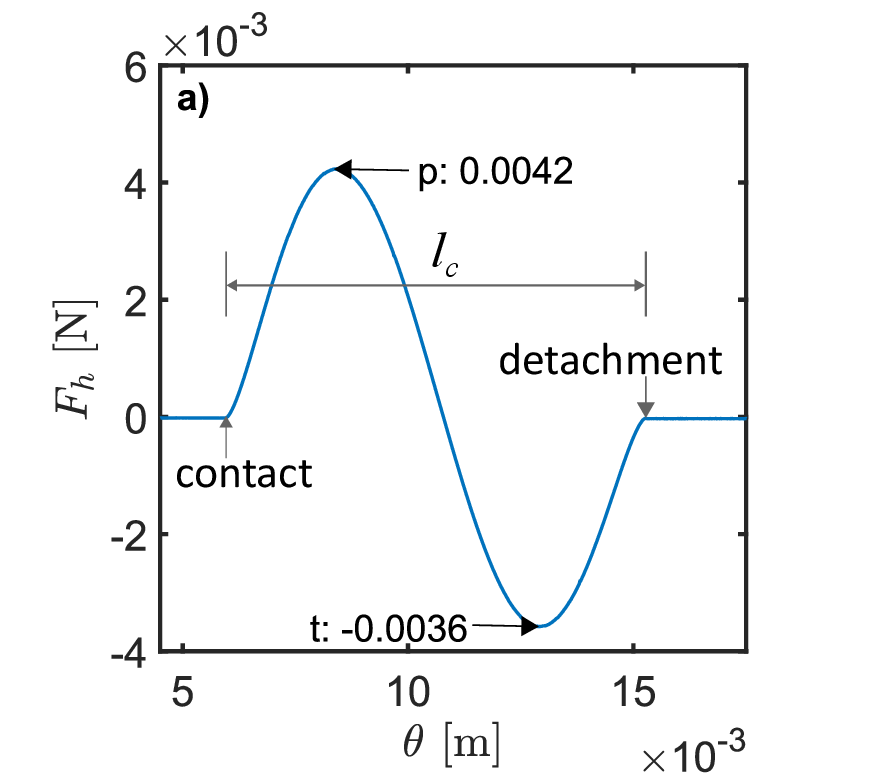} 
    \includegraphics[scale=0.4]{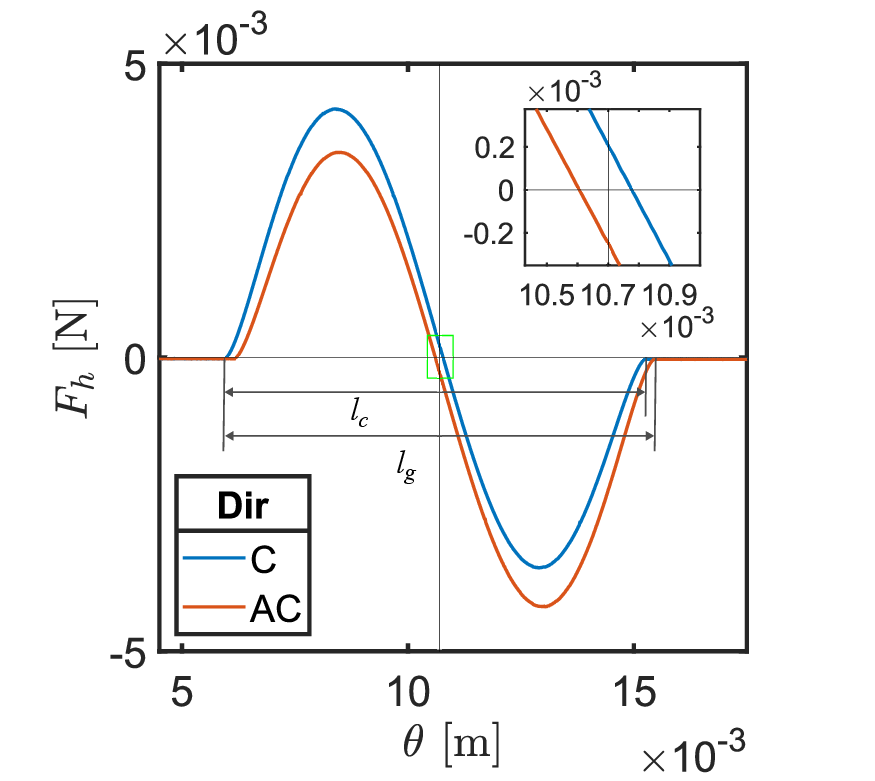}
    \caption{ a) Horizontal forces $(F_v)$ at constant maximum $\delta_{max}$ overlap and at constant rotational velocity $(v)$ as the probe moves clockwise b) main: Horizontal forces $(F_v)$ at constant maximum $\delta_{max}$ overlap and at constant rotational velocity $(v)$ as the probe moves clockwise (C: in blue) and anticlockwise (AC: in red); inset: force response at the center of geometric contact length (original area highlighted in green) }
    \label{fig:Hz_force_const_del_v}
\end{figure*}

\paragraph*{Constant central overlap and velocity}
We first examine the horizontal force response from the interparticle shear experiment in a fixed gap and at a constant velocity, shown in figure \ref{fig:Hz_force_const_del_v}a). In this figure, the upper particle is moved in the clockwise direction. Figure {\ref{fig:Hz_force_const_del_v}}a) plots the change in force in the horizontal plane ($F_h$) as a function of linear angular distance travelled ($\theta$) from the rheometer's zero deflection angle reference point. Here, the increase in $F_h$ on the left marks the formation of a contact. The detachment is marked by the return of the forces back to a constant baseline. We can so establish a total distance of contact as the difference between the $\theta$ value at the point of contact and the $\theta$ value at the point of detachment. We refer to this distance as the deformed contact length ($l_c$). This length, along with the points of contact and detachment are visualized in figure {\ref{fig:Hz_force_const_del_v}}a). We call $l_c$ the deformed contact length, as the length determination is affected by the strain on the hydrogel sphere and the sliding velocity. The figure also clearly shows that the peak magnitude of the force response (indicated with $\leftarrow$p), as the moving particle approaches the stationary particle's center, is larger than the trough magnitude (indicated with t$\rightarrow$) as the moving particle moves away from the stationary particle's center. Thus, an asymmetry in the force response in the horizontal direction is observed. 

After the deformed contact length determination in one rotation direction, we can rotate in the opposite direction to establish the true contact length. Figure {\ref{fig:Hz_force_const_del_v}}b) shows the horizontal force responses ($F_h$) when the test is repeated with the upper particle being moved in the opposite direction (the anticlockwise (AC) direction) in red, along with results from figure {\ref{fig:Hz_force_const_del_v}}a) where the upper particle was moved clockwise (C) in blue.  The anticlockwise (AC) experiment measure the undeformed contact point of the hydrogel, which is needed to establish the undeformed contact length, the we call undeformed contact length (or the geometric contact length if the particle were perfectly rigid) $l_g$ . It needs to be kept in mind that the point of contact for the blue curve is on the left of the curve and the point of contact for the curve in red is on the right side. To clarify why we need two length definitions, we can take a closer look at the data in figure {\ref{fig:Hz_force_const_del_v}}. Here, we notice that the point of detachment as the rotating arm is moved clockwise direction (blue data) does not spatially coincide with the position of the point of contact detected as the rotating arm is moved in the anticlockwise (AC) direction. The point of detachment in AC direction is similarly different from the point of contact in the C direction. These observations mean that $l_g$ is longer than $l_c$ in both the clockwise and anticlockwise direction. For context, the difference between the deformed contact lengths $l_c$ in the clockwise and anticlockwise direction was only 5.5 $\mu m$ whereas the difference between the average contact length $l_c$ and geometric length $l_g$ was 219.2 $\mu m$ for the dataset shown in this figure.

Furthermore, if we zoom in to the center of the geometric contact in figure \ref{fig:Hz_force_const_del_v}b) inset, it can be noticed that the zero crossing in the horizontal forces (the point where the horizontal forces change from positive to negative or vice versa) is not at the center of the contact distance. The zero crossing in fact moves away from the point of contact towards the point of detachment in both directions. This implies that there is a finite but small horizontal force acting on the particles even when they are directly on top of each other.

\begin{figure*}
    \centering
    \includegraphics[scale=0.4]{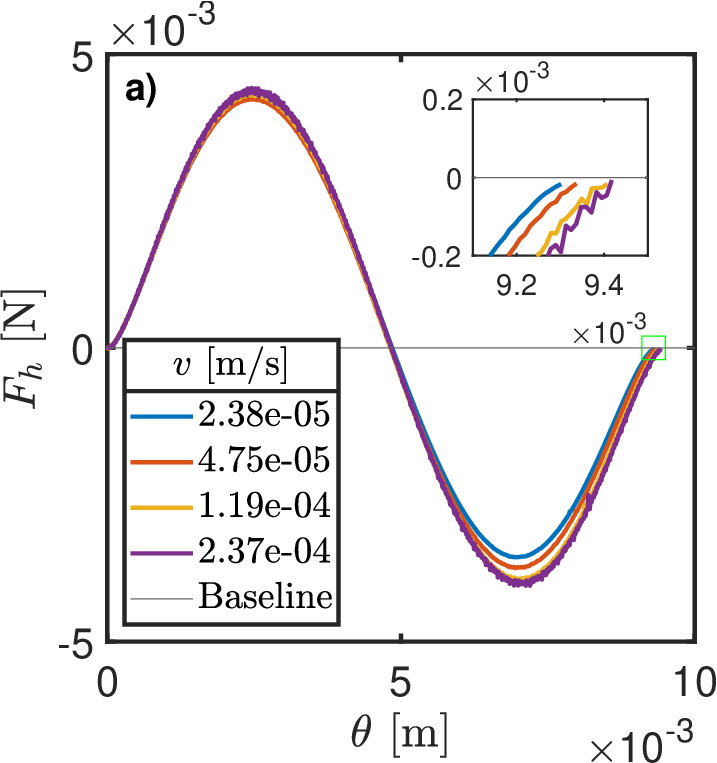}
    \includegraphics[scale=0.4]{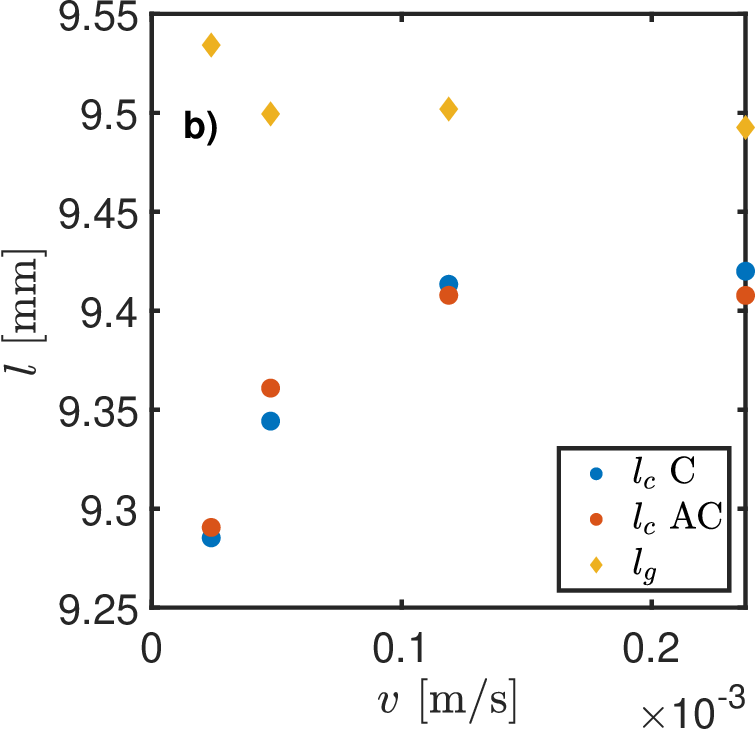}
    \caption{a) main: Horizontal Force $(F_h)$ response at a constant maximum overlap ($\delta_{max}$ = 0.78 mm and mean particle diameters of 19.1 mm and 18.4 mm, maximum strain ($\epsilon_{max}) =0.0208$) but at different velocities $(v) m/s$ in the clockwise direction; inset: force response towards the end of geometric contact length (original area highlighted in green)  b) change in contact lengths ($l mm$) as a function of rotational velocity ($v m/s$) where, geometric contact length are $l_g$  (in yellow) and the deformed contact lengths are $l_c$ and the directions are specified as C (clockwise) and AC (anticlockwise) at a a constant maximum overlap ($\delta_{max}$ = 0.78 mm)}
    \label{fig:Hz_force_const_del_varaiable_v}
\end{figure*}

\paragraph*{Velocity dependence of the contact force}\label{constant_overlap_variable_vel}
Figure \ref{fig:Hz_force_const_del_varaiable_v}a) compares the horizontal force response of spherical hydrogels in response to sliding at different constant velocity levels. The first feature that is apparent from this figure is that the amplitude of the troughs for the different sliding velocities do not coincide. On closer inspection, we notice that the amplitude of these troughs is greater for larger velocities. The same is also true for the peaks, but this feature is less pronounced in the plot as the relaxation is contact duration dependent, and at the peaks, the contact is younger. Thus, a clear velocity-dependent response can be observed. When we recall equation \ref{eq:modulus_division_expanded} which essentially models the effective shear modulus as a decaying function with time, this experimental model makes sense. Experiments done at lower sliding velocities take a longer time and therefore work with a lower effective shear modulus; thus their force response to the same strain or overlap is lower. 

Another feature of interest can be seen at the very end of the contact. The force response just before the two particles detach is highlighted in the inset in figure \ref{fig:Hz_force_const_del_varaiable_v}a). Here we see that even though all of these tests were performed on the same set of particles, the points of detachment at different velocities do not coincide. The particle seems to detach earlier when the particles move slower relative to each other. This too is consistent with the model explained in equation \ref{eq:modulus_division_expanded}. In figure \ref{fig:Hz_force_const_del_varaiable_v} b) we plot the the geometric contact length ($l_g$) and the deformed contact length ($l_c$) obtained at the different sliding speeds ($v$) at which the experiments were performed. Here, we clearly see that the variation in $l_g$ at different speeds is much lower than the variation of $l_c$. Furthermore, we see that $l_c$ seems to grow at larger sliding speeds and approaches $l_g$. This trend is consistent when the experiment is performed in either direction as can also be seen in the figure \ref{fig:Hz_force_const_del_varaiable_v}.

\paragraph*{Recreating experimental features with the analytical model}

\begin{figure*}
    \centering
    \includegraphics[scale=0.56]{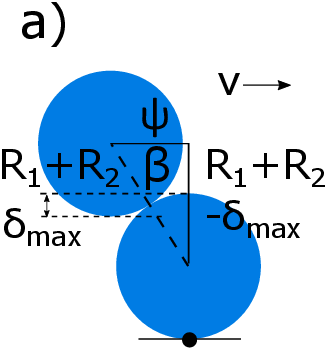}
    \includegraphics[scale=0.5]{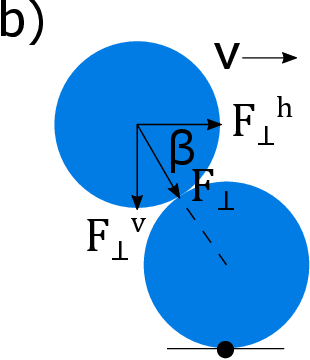}
    \includegraphics[scale=0.5]{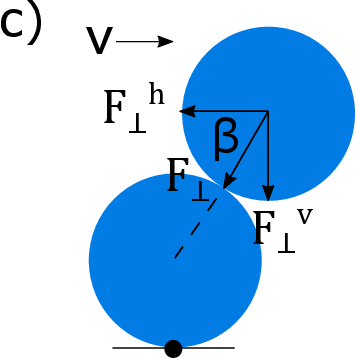}
    \includegraphics[scale=0.5]{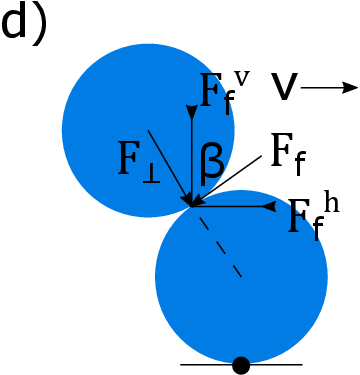}
    \includegraphics[scale=0.5]{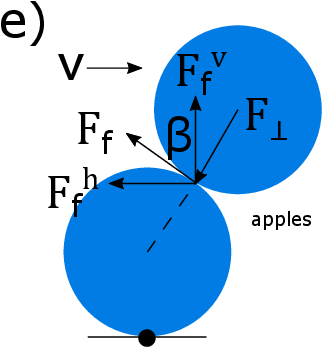}
    \caption{Resolution of force components a) Determination of angle between horizontal and center to center force vectors ($\beta$) b) Resolution of elastic and dissipative forces (acting between particle centers) as the moving particle approaches the stationary particle c) Resolution of elastic and dissipative forces (acting between particle centers) as the moving particle moves away from the stationary particle d)Resolution of friction forces (acting tangentially) as the moving particle approaches the stationary particle e)Resolution of friction forces (acting tangentially) as the moving particle moves away from the stationary particle}
    \label{fig:vector_diagrams}
\end{figure*}

Figure \ref{fig:vector_diagrams} describes how the forces from the model described in section \ref{theoretical_model} are vectorially resolved so that they can be compared to the data observed in the experiment. The two dimensional analytical model consists of measuring the effective overlap of two circles moving past each other. The upper circle is moving from left to right at a velocity $v$ and the lower circle is fixed. Figure {\ref{fig:vector_diagrams}}a) shows the geometric quantities involved in calculating the total forces in the horizontal and vertical directions at contact. At contact, the distance between the particle centers are $R_1+R_2$. $\delta_{max}$ is the the maximum overlap between the particles. Thus the vertical distance between the particle centers is $R_1+R_2-\delta_{max}$. $\beta$ is the angle between the horizontal plane and the line joining the two particle centers and can be calculated using the triangle shown in figure {\ref{fig:vector_diagrams}}a). $\psi$ is the horizontal distance between the particle centers which can also be calculated from the same figure, at the moment of contact $\psi=\psi_0$. After this moment (at $t=0$), the overlap ($\delta$) is zero. After this moment, $t$ grows from 0 and $\psi=\psi_0-vt$. The vertical distance between the particle centers remains $R_1+R_2-\delta_{max}$ and a new distance between particle centers can be calculated as $\sqrt{(R_1+R_2-\delta_{max})+(\psi_0-vt)}$. These quantities are also used to calculate a new $\beta$. Similarly, $\delta=(R_1+R_2)-\sqrt{(R_1+R_2-\delta_{max})+(\psi_0-vt)}$.), for any instant of time $t$, when the horizontal displacement $\theta=vt$.

These geometric quantities are then used to calculate the $F_{\perp}(t)$ values using the equations described in section {\ref{theoretical_model}}. The values of the shear moduli $G_e$, $G_l$ and timescale $\tau$ were manually fit to the experimental vertical and horizontal forces independently. The values reported here are the ones that fit the experimental vertical and horizontal forces best at all the velocities and overlaps explored in this work. Figure {\ref{fig:vector_diagrams}}b) shows the vector resolution of $F_{\perp}$ into its horizontal component $F_{\perp}^h$ and vertical component $F_{\perp}^v$ when the upper circle approaches the fixed lower circle from the left. Figure {\ref{fig:vector_diagrams}}c) shows the vector resolution of $F_{\perp}$ when the upper circles moves away from the pinned lower circle. It can be noted that for undeformable circles as assumed in his diagram, the direction of the vertical components remain the same regardless of whether the moving circle approaches or moves away from the pinned circle whereas the direction of the horizontal components reverses. As we will see, the elastic deformation of the spheres will induce an asymmetry in the force response, to which we will return below in the consideration of contact length determinations.

Figure {\ref{fig:vector_diagrams}}d) shows the vector resolution of $F_f$, which is the frictional force between the circles calculated using equation {\ref{eqn:friction}} into its horizontal component $F_f^h$ and vertical component $F_f^v$ when the upper circle approaches the pinned lower circle from the left. Figure {\ref{fig:vector_diagrams}}e) describes the vector resolution of $F_f$ as the upper moving particle moves away from the pinned circle. Here, the direction of the vertical component of the force $F_f$ reverses when the moving circle moves away from the pinned circle compared to when the moving circle approaches the pinned circle. This reversal, however will be visible in the resultant vertical force as the magnitudes of  $F_f$ is linked to $F_{\perp}$, and $F_{\perp}$ will be small towards the end of the contact. The direction of the horizontal components, however remains the same and leads to the offset in the $F_hz$ at the center of the contact as shown in the inset of figure {\ref{fig:Hz_force_const_del_v}}b.

\begin{figure*}
    \centering
    \includegraphics[scale=0.41]{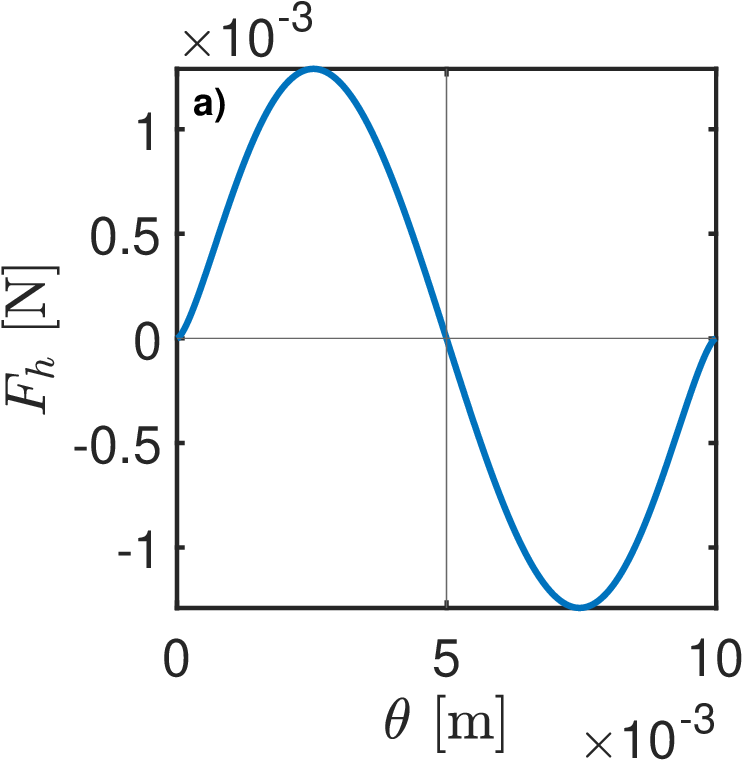}
    \includegraphics[scale=0.41]{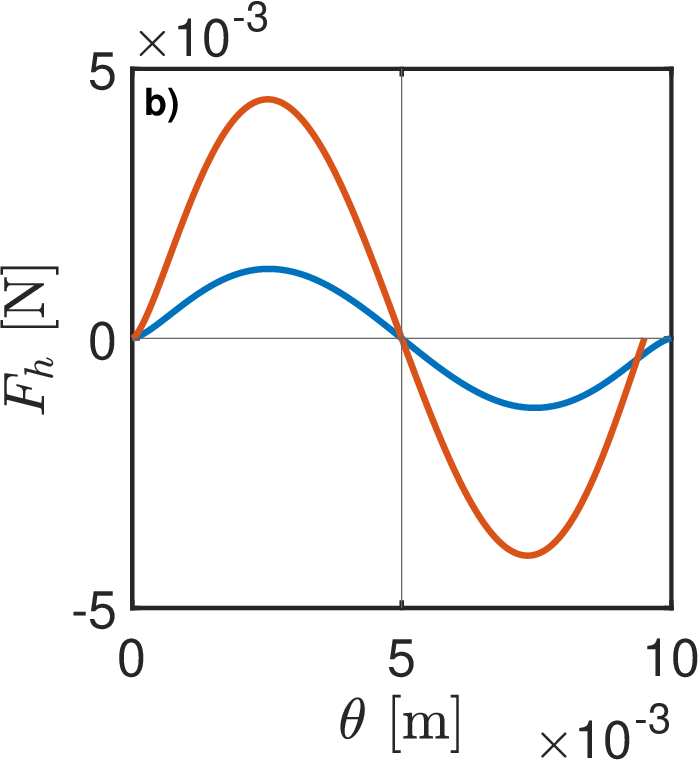}
    \includegraphics[scale=0.41]{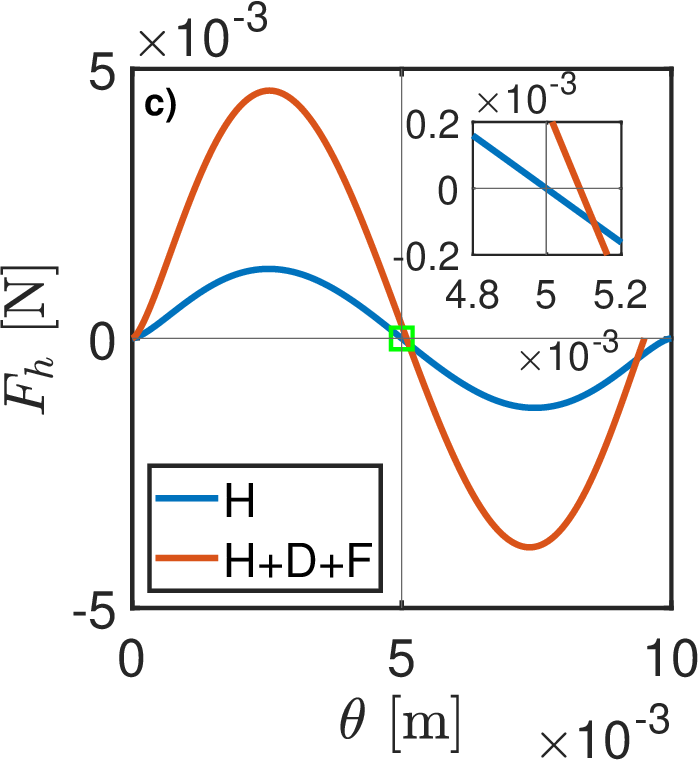}
    \caption{$F_h$ responses from the model after vector resolution from a) the elastic Hertz model (H) alone  b) the elastic model (H) added to the Jian's dissipative model (D), with the cutoff eliminating adhesive forces implemented in it c)The elastic (H) and dissipative (D) force response coupled to the Coulomb's friction model (F). Here, the coefficient of friction ($\mu$) used is $5\times10^{-3}$}
    \label{fig: model_responses}
\end{figure*}

Figure \ref{fig: model_responses} describes how each element of the model adds up to explain the various features we see in the experiment. Figure \ref{fig: model_responses}a) simply shows the sine-like change in $F_h$ as a function of the linear angular displacement ($\theta$=angular displacement[rad]$\times$ moment arm $0.027$ [m]) after the vector resolution of the elastic force with changing $\beta$ and overlap $\delta$. In Figure \ref{fig: model_responses}b) we add the viscolelastic response from Jian's model described in section \ref{viscous} to introduce the velocity dependence we observe in section \ref{constant_overlap_variable_vel}. However using the model as is, induces an adhesive force towards the end of the contact. This does not match our experimental results as we see no adhesion. Thus, the cutoff described in section \ref{viscous} is implemented so as to remove any adhesive forces. This gives us both a velocity dependent force response and a velocity dependent contact length $l_c$. Here, it is interesting to note that the velocity dependent $l_c$ results as a byproduct of the integration scheme and the adhesion cutoff. The model itself still assumes that the overlap between the particles is positive in this region and that the rate of change of overlap is negative. Therefore, switching to a different viscous force law cannot give us a zero force response when both the overlap ($\delta$) and the rate of change of overlap ($\dot{\delta}$) are non-zero. 
However, even with these models the zero crossing in the $F_h$ still remains at the center of the geometric contact length $l_g$. If we look at the inset in figure \ref{fig:Hz_force_const_del_v}b), this is not the case in our experiments, meaning that $F_h$ is not zero when the sliding particles are on top of each other. Therefore, the center to center force responses from the elastic and dissipative force laws are coupled with a friction model, which allows us to model a non zero horizontal force ($F_h$) at the center of the geometric contact length $l_g$. This can be seen in figure \ref{fig: model_responses} c), especially in the inset where the $F_h$ at zero crossing has been zoomed into. For reference, the elastic Hertzian response is plotted in blue and we see that this response passes through the center of $l_g$.

\begin{figure*}
    \centering
   \includegraphics[scale=0.32]{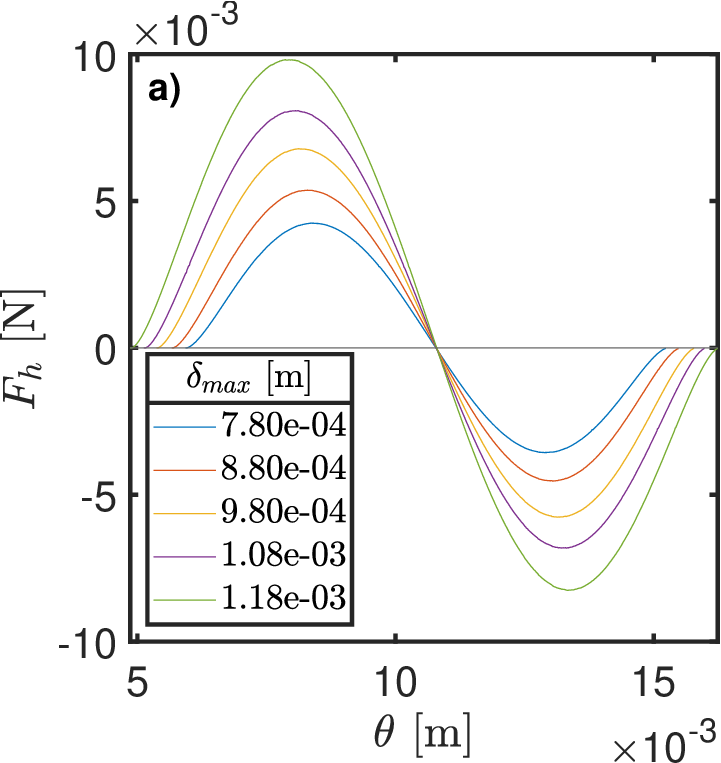}
   \includegraphics[scale=0.32]{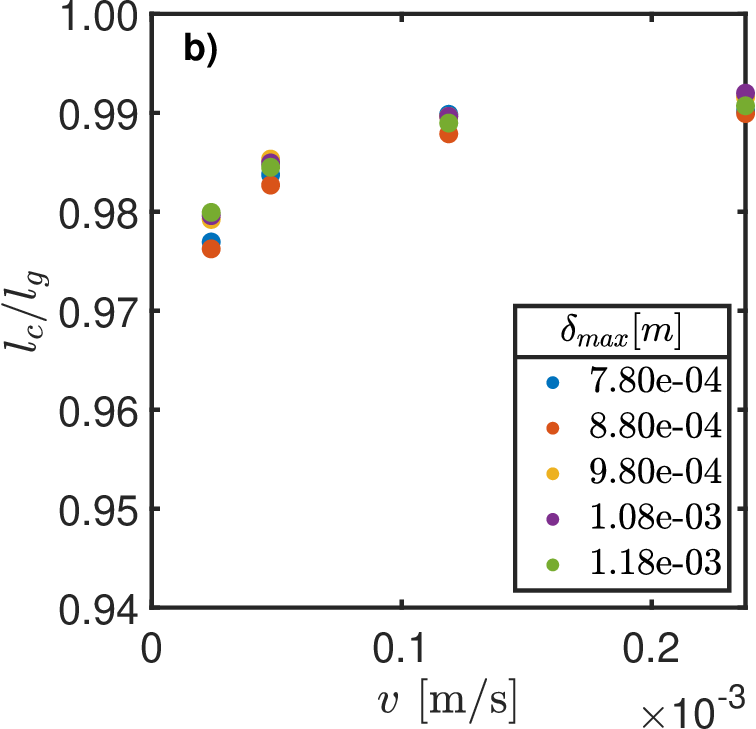}
   \includegraphics[scale=0.32]{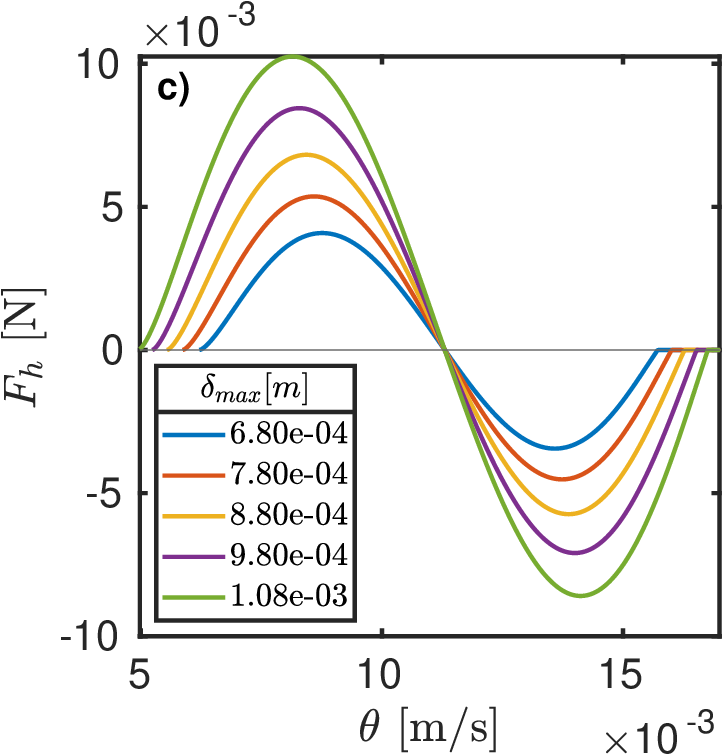}
   \includegraphics[scale=0.32]{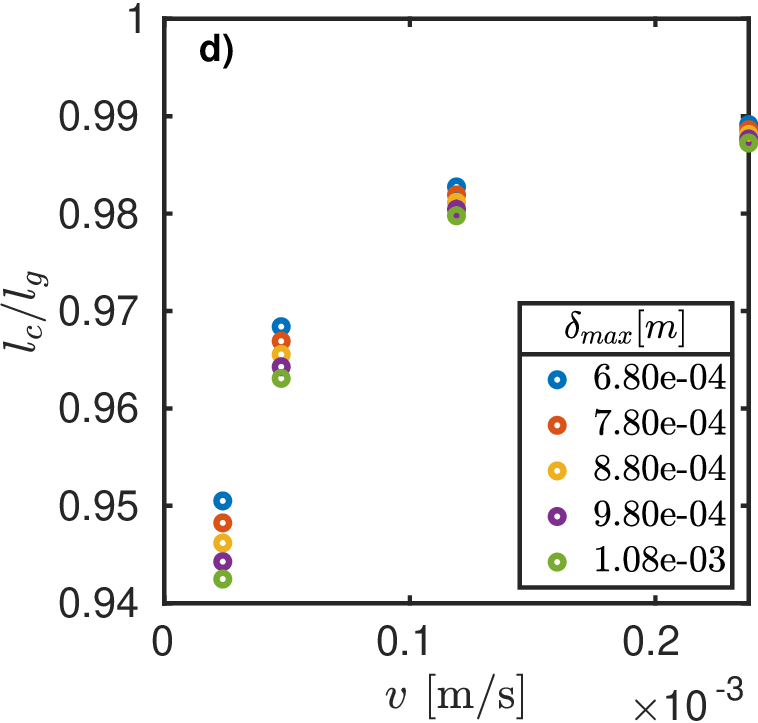}
    \caption{Results from the experiment: a) Horizontal force $(F_h)$ response at a different maximum overlaps ($\delta_{max} m$) at the sliding speed $(v)=2.38\times10^{-5}$ m/s in the clockwise direction vs. the distance at which this force was observed from the reference point b) change in the ratio of deformed contact length ($l_c$) and geometric contact length ($l_g$) as a function of sliding velocities ($v$) for experiments done on the same particle pairs at different maximum overlaps ($\delta_{max}$); Results from the numerics: c) Horizontal force $(F_h)$ response at a different maximum overlaps ($\delta_{max} m$) at the sliding speed $(v)=2.38\times10^{-5}$ m/s in the clockwise direction vs. the distance at which this force was observed from the reference point d) change in the ratio of deformed contact length ($l_c$) and geometric contact length ($l_g$) as a function of sliding velocities ($v$) for experiments done on the same particle pairs at different maximum overlaps ($\delta_{max} m$)}
    \label{fig:exp_hz_data_analyis_multiple_overlap}
\end{figure*}

\paragraph*{Robustness with increased particle overlap}
Figure \ref{fig:exp_hz_data_analyis_multiple_overlap}a) shows the horizontal force response of five interparticle shear tests carried out in incremental amounts of maximum overlap ($\delta_{max}$) at a sliding speed of $2.38\times10^{-5}$ m/s in the clockwise direction. Similar data were also measured for sliding speeds for all four speed levels and in the anticlockwise direction as in the previous experiments. In this figure, we see that both the contact lengths and the amplitude of horizontal force responses grow as a function of maximum overlap. It is also notable to see that the point where the sign of the amplitude flips for all velocities almost coincides. This shows that the particles did not appreciably move in the horizontal direction between these experiments: the shell and collar hold the particles firmly in place. 

The results obtained in figure \ref{fig:exp_hz_data_analyis_multiple_overlap}a) by themselves are expected, as the force response increases when the overlap or strain between two particles is increased. However, given that we saw that the geometric contact length and the actual length are different and velocity dependent in figure \ref{fig:Hz_force_const_del_varaiable_v}b), the results of how these might scale with an increase in overlap is interesting. Since the contact lengths obviously must scale with the overlap, a direct comparison is difficult. Thus we normalize $l_c$ by $l_g$ (since $l_g$ is a geometrical parameter that scales with particle radii and the $\delta_{max}$). Thus $\frac{l_c}{l_g}$ is plotted against the sliding speed in figure \ref{fig:exp_hz_data_analyis_multiple_overlap}b). The plot reveals that the average deformed contact length ($l_c$) does indeed scale with sliding speed ($v$) and grows towards $l_g$ as $v$ increases.

The results seen in the experiments can also be qualitatively reproduced using the model described in section \ref{theoretical_model} as can be seen in figures \ref{fig:exp_hz_data_analyis_multiple_overlap}c) and d). These figures can be compared directly to the experimental results in figures \ref{fig:exp_hz_data_analyis_multiple_overlap}a) and b) respectively. However, since the particles used in this experiment are not perfectly round there are deviations in the radii of particles and maximum overlap used in the experiment and in the model. These are clearly visible from the difference in overlaps shown in the experimental results and the results from the model. The parameters for obtaining these results were: $G_0=0.4 \times 10^4 Pa$, $G_l=5 \times G_0$, $\tau = 2000 sec$.

\subsubsection{\label{experimental_ver_forces} $F_v$ at constant overlap}
\begin{figure*}
    \centering
    \includegraphics[scale=0.4]{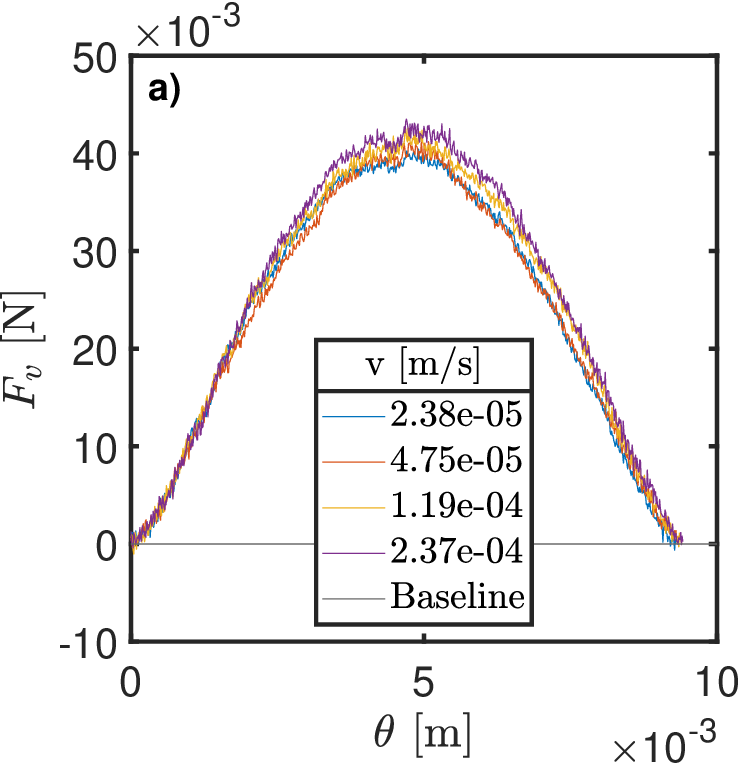}
    \includegraphics[scale=0.4]{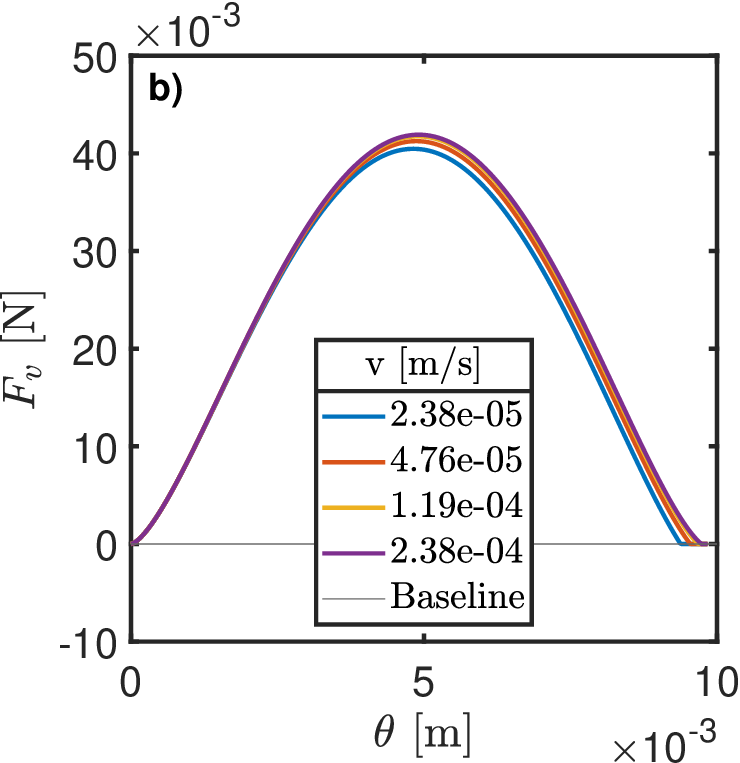}
    \includegraphics[scale=0.4]{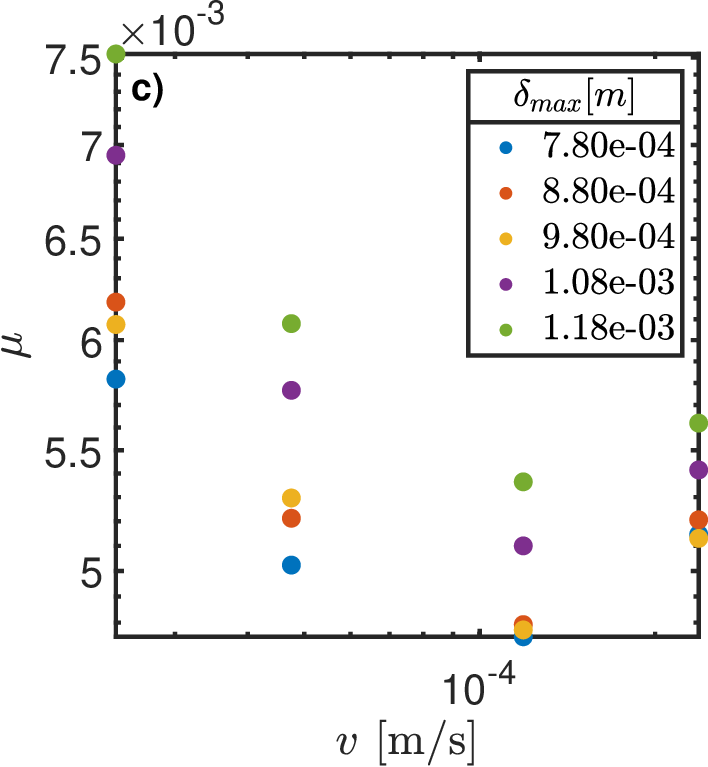}
    \caption{Vertical component of forces $(F_v)$ during a shear experiment at different sliding velocities $(v)$ in a) the experiment b) the model c) Variation in coefficients of friction ($\mu$) as a function of sliding velocity ($v$) for different overlaps}
    \label{fig:vertical_forces}
\end{figure*}

Figure \ref{fig:vertical_forces}a) shows the vertical component of the force between the particles as a function of probe displacement for a sliding spherical hydrogel at the same maximum experimental overlap. From this figure, we see that the normal force responses for the different sliding velocities at a constant overlap are slightly velocity dependent. However, the velocity dependence is difficult to distinguish because of the noise level in the normal force data. The determination of $l_g$ is similarly affected by the noise, and hence not shown.

Figure \ref{fig:vertical_forces}b) shows the numerically obtained vertical component of the forces using the combination of models described in section \ref{theoretical_model} and is able to reproduce similar velocity-dependent trends as shown by the experimental data in figure \ref{fig:vertical_forces}a). The parameters used to obtain these results were: $G_0=0.4 \times 10^4 Pa$, $G_l=4 \times G_0$, $\tau = 2000 sec$. The values of $G_0$, $G_l$ and $\tau$  are obtained from a manual fit procedure to best match the experimental results shown in figure 8a). Here, the $G_l$ for the vertical force component that resulted in the best amplitude match to the experimental results was 4/5ths of the $G_l$ for the horizontal force components. We believe that this anisotropy in $G_l$ in the horizontal and vertical force components is due to the difference in geometrical constraints in the two force directions. This is because, when the upper particle slides over the fixed particle, the horizontal component of the compressive forces will make the hydrogels expand in the vertical direction, which is possible because the top of the gels are largely unrestrained.  Similarly, the vertical component of the compressive force will try and make the hydrogels expand in the lateral direction. This will not be possible because the gel is restrained by the shell and collar assembly, creating a confinement. This induces a confining stress that limits the viscous response from the hydrogels.

Additionally, we can extract the horizontal forces at the center of the geometric contact length ($l_g$) shown in figure \ref{fig:Hz_force_const_del_v}c), and the vertical forces at the center of the geometric contact length ($l_g$) by fitting a second-degree polynomial to 20 data points left and right of the center of ($l_g$). The ratio of these forces gives us a way to calculate the coefficient of friction ($\mu$) at the point where we know the overlap ($\delta$) and where the vertical and horizontal force theoretically align with the normal and tangential components of forces. Figure \ref{fig:vertical_forces}b) plots the $\mu$ obtained against the sliding velocities for the different overlaps tested. While a master curve is not obtained, the plot does show how the $\mu$ changes at different velocities. This is reasonable because the all the tests shown in this study constitute of lubricated contacts as the hydrogels used make contact underwater, and are also poroelastic. The frictional force in lubricated contacts can be a function of normal load, viscosity and velocity in contacts, as already experimentally shown by Stribect et. al. \cite{stribeck1901kugellager}. A more complete treatment of the hydrogel-hydrogel friction is outside the scope of this work, but can be found in other works, e.g. \cite{gong2006friction, pitenis2014polymer, workamp2019contact}. 

\subsection{\label{Timescale}Plate-plate compression}
\begin{figure*}
    \centering
    \includegraphics[scale=0.4]{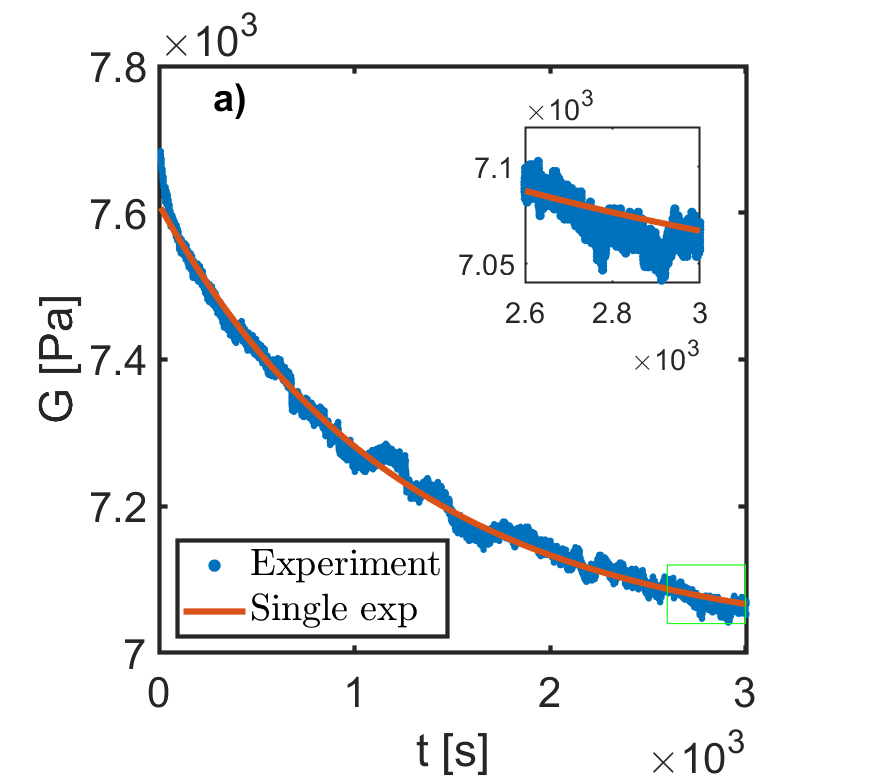}
    \includegraphics[scale=0.4]{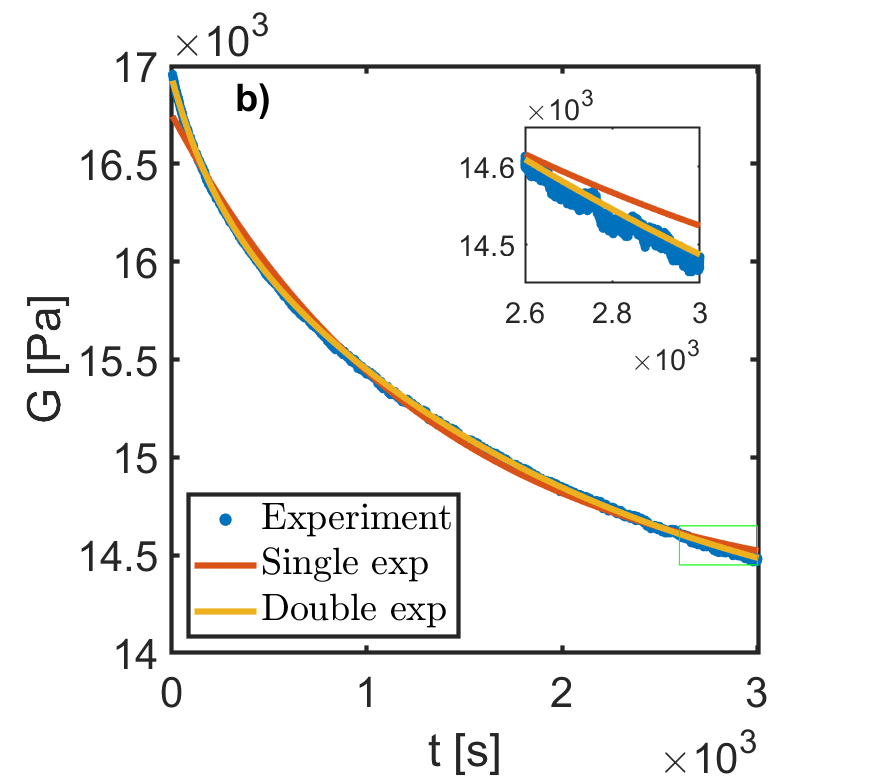}
    \caption{a) Decay in elastic moduli vs. time when the hydrogel is compressed between two plates: blue points are the calculated elastic moduli assuming perfect Hertzian contact and their supposed evolution with time, the red line is fit with equation \ref{eq:modulus_division_expanded} (in red), inset: force response towards the end of the the relaxation experiment b) Decay in elastic moduli vs. time when the hydrogel is placed in a 3D printed shell described in section \ref{shell} and compressed with a plate: blue points are the calculated elastic moduli assuming perfect Hertzian contact and their supposed evolution with time; inset: force response towards the end of the the relaxation experiment; the red line is fit with equation \ref{eq:modulus_division_expanded} (in red), and the yellow line an additional fit with a second exponential time to better account for the systematic deviation from the single exponent fit at the beginning and end of the experiment that is more obviously visible in the inset}
    \label{fig:Elastic_modulus_decaymatlab}
\end{figure*}

Section {\ref{interparticle_results}} establishes a timescale in the force response in the tangential and center to center direction during a rearrangement event. In this section, we show that a similar force response can be found when the particle is under a fixed level of vertical compressive strain. We already see that the vertical force response decays in time during the sliding test from figure {\ref{fig:setup_compression}}c). We use the vertical force evolution with time dataset similar to figure {\ref{fig:setup_compression}}c), at a constant $\delta$ and back calculate the elastic moduli of the material at each time step using equations {\ref{eq:elastic_equation_force}} to {\ref{eq:elastic_equation}} (assuming a linear elastic model). Thus, we get a shear modulus $G$ at each time step. Figure {\ref{fig:Elastic_modulus_decaymatlab}}a) shows results for this shear modulus ($G$) as a function of time for a particle that is compressed between two flat plates.  Figure {\ref{fig:Elastic_modulus_decaymatlab}}b) plots the extracted shear modulus($G$) as a function of time ($t$) when the same hydrogel sphere is placed in the custom 3D printed geometry described in section {\ref{shell}} that was also used for the inter-particle shear experiments. As can be seen by comparing figures {\ref{fig:Elastic_modulus_decaymatlab}}a) and b), this leads to a difference in the calculated $G$ values, even though the sphere being tested is the same. For the experiments in which the hydrogel sphere is confined to the 3D printed geometry we interpret the difference from the fact that the sphere could not extend laterally and had a larger contact with the 3D printed geometry.

The back calculated shear modulus $G$ vs. time $t$ plots in figure {\ref{fig:Elastic_modulus_decaymatlab}} are then fitted to a decaying exponential function shown in equation {\ref{eq:modulus_division_expanded}}. These fits are shown in red. For figure {\ref{fig:Elastic_modulus_decaymatlab}}a) this results in a $G_e$ of 7146 Pa, $G_l$ of 630 Pa and a $\tau$ of 1413 sec. In figure {\ref{fig:Elastic_modulus_decaymatlab}}b), for a hydrogel sphere placed in a 3D printed custom geometry compressed by a flat plate, we see that a single timescale is inadequate to capture the response. The single exponent fit described by equation {\ref{eq:modulus_division_expanded}} deviates from the data both at the start of dataset and the end of the dataset. Thus, a second exponent was added to equation {\ref{eq:modulus_division_expanded}} to fit the entire data series changing it to $\hat{G}(t)=G_e + G_{l1}e^{(-\frac{t}{\tau_1})} + G_{l2}e^{(-\frac{t}{\tau_2})}$. The fit resulted in a $G_e$ of 13915 Pa, $G_{l1}$ of 2540 Pa, $\tau_1$ of 1898 sec, $G_{l2}$ of 504 Pa and $\tau_2$ of 172 sec. Here, $G_{l1}>G_{l2}$ so the timescale being considered is 1898 sec. Therefore, the experiments together establish a consistent timescale of response at the particle scale of about 1500 seconds, which is similar to the timescale used to model the shear experiments in section {\ref{interparticle_results}}. 

\section{Discussion}

Our results show that contact mechanics between soft, almost frictionless grains even in a single contact, are affected by a combination of the grain's volumetric and surface properties and its geometry. Both the inter particle shear tests and the plate plate compression tests reveal that there is a long relaxation timescale. We believe the timescale to be associated with the grain material property. The second, faster timescale seen in the plate-plate compression tests, is not explored in the work here, as the shear experiments were done at a very slow speed, making the short relaxation mode unobservable in the sliding force dynamics. It must also be noted that the velocity and overlap ranges explored in this experiment are limited, which means that more timescales, and potentially a whole spectrum of modes, are potentially present in the hydrogel (contact) mechanics as can be observed in other studies ({\cite{malkin2001dynamic,xu2013understanding}}). The relaxation already observed may be playing a role in the observed slow (creep) dynamics in soft particle packings: slow contact force evolution affects force balance and can so induce particle motion.\cite{dijksman2022creep}.


\section{\label{Conclusion}Conclusion}
We constructed a custom rheometric tool to measure the contact forces between two submersed hydrogels sliding past each other at different overlap amplitudes and different sliding velocities. We also measure the compressive force response of single hydrogel beads. We use Jian's dissipative model \cite{jian2019normal} to rationalize the observations on center-to-center forces between particles during shear and couple it with tangential forces using a simple Coulomb friction. We are so able to describe our experimental observations with five parameters; the constant elastic part of the modulus, the maximum relaxable part of the shear modulus, a single relaxation time, a Poisson's ratio, and a coefficient of friction. By qualitatively matching the data from the model to those of the experiments, we can show that these five constants can describe forces evolving in both center-to-center and tangential directions. Our results establish a time-dependent contact dynamics framework for single-hydrogel particle contacts that lets the material properties of the contacting particles evolve from the moment of contact. The results of our experimental work and numerical validation show that in (numerical, theoretical) soft particle contact mechanics, not only the particle position dynamics should be taken into account in the evolution of the contact forces and the network structure, but also relaxation effects of the bulk material out of which the particles are composed. Further work will concentrate on how these moduli propagate into bulk properties. We will also investigate whether the framework described above can indeed be used to obtain a microscopic understanding of properties such as creep in hydrogel packings, as was observed by \cite{dijksman2022creep}.

 






\section*{Conflict of Interest Statement}

The authors declare that the research was conducted in the absence of commercial or financial relationships that could be construed as a potential conflict of interest.

\section*{Funding}
This project has received funding from the European Union’s Horizon 2020 research and innovation programme under the Marie Skłodowska Curie grant agreement number 812638.

\section*{Acknowledgments}
We would like to acknowledge the efforts of the team at PCC at Wageningen University. In particular, we thank our technicians Raoul Fix and Remco Fokkink, without whom the setups on which these experiments were done could not be built. We would like to thank our colleagues Jose Ruiz Franco and Lawrence Honaker for their helpful insight regarding this project. We would also like to thank the team at TU Graz and DCS including but not limited to Stefan Radl, Fransisco Goio Castro, and Michael Mascara for the lively discussions on this topic. We would also like to thank Pinzon from Labortoire 3SR, Grenoble, for pointing to older literature for similar works done in this sector with non-deformable particles. Special thanks to Brian Tighe for pointing us in this direction.

\section*{Data Availability Statement}
All the data associated with this paper is available upon request and is publicly available in zenodo using \cite{shakya_2024_10824828}.


\bibliography{apssamp}

\end{document}